\documentclass[aps,prl,twocolumn,10pt,superscriptaddress,floatfix]{revtex4-2}

\usepackage{amsmath,amssymb,amsfonts}
\usepackage{graphicx}
\usepackage{dcolumn}
\usepackage{bm}
\usepackage[hidelinks]{hyperref}
\usepackage{xcolor}

\newcommand{\vF}{v_{\mathrm{F}}}
\newcommand{\hbarvF}{\hbar\vF}
\newcommand{\rR}{r_{\mathrm{R}}}
\newcommand{\rL}{r_{\mathrm{L}}}

\newcommand{\Rm}{R_m}
\newcommand{\sgn}{\operatorname{sgn}}
\begin{document}

\title{Dirac Fermion Scattering and Conductance Response in Asymmetric Graphene Wormholes}

\author{Arian~Gorza}
\email{arian.gorza@uns.edu.ar}
\affiliation{Departamento de F\'isica, Universidad Nacional del Sur and
       CONICET, Av.\ Alem 1253, B8000CPB Bah\'ia Blanca, Argentina}

\author{Facundo~Arreyes}
\affiliation{Departamento de F\'isica, Universidad Nacional del Sur and
       CONICET, Av.\ Alem 1253, B8000CPB Bah\'ia Blanca, Argentina}

\author{J.\,S.~Ardenghi}
\affiliation{Departamento de F\'isica, Universidad Nacional del Sur and
       CONICET, Av.\ Alem 1253, B8000CPB Bah\'ia Blanca, Argentina}

\begin{abstract}
We calculate massless Dirac scattering and the conductance response to a localized scalar ring gate in a graphene wormhole with two independent curvature scales and no axial magnetic flux. Matching Hankel and Gauss hypergeometric spinors gives the unperturbed scattering states. Interchanging the throat segments preserves the total Landauer conductance, while a gate at a fixed displaced position produces different conductance changes in the two mirror configurations. For $a=5$~nm and $r_L+r_R=13.68$~nm, a $1$~meV gate of width $1$~nm centered at $u_0=7$~nm gives relative conductance changes of $+0.0718\%$ and $-0.0918\%$ at $E=200$~meV for $\eta=0.5$ and $2$, respectively. Energy--position maps show how asymmetry redistributes the electrical response. Gates centered well inside both curved segments can give different response magnitudes; opposite signs occur near the mouths and farther out. We also find opposite response signs in the fixed-length comparison at the chosen operating point. For equally occupied opposite angular channels, the sublattice imbalances cancel and the conductance responses add. A spatially controlled gate thus probes throat orientation through conductance even when the unperturbed two-terminal measurement cannot distinguish the two configurations.

\end{abstract}

\maketitle
\raggedbottom

\section{\label{sec:intro}Introduction}
Low-energy charge carriers in graphene behave as $(2{+}1)$-dimensional massless Dirac fermions~\cite{Novoselov2004,CastroNeto2009,DasSarma2011}, with amplitudes on the $A$ and $B$ sublattices forming a two-component pseudospinor; local perturbations can modify the spectrum and sublattice weights~\cite{Uchoa2009,Kotov2012}, although unequal sublattice weights are also compatible with a gapless spectrum.

Planar graphene preserves spatial inversion symmetry; however, macroscopic out-of-plane deformations modify the local electronic structure~\cite{Vozmediano2010,Cortijo2007}. 
The coupling between geometry and the charge carriers is introduced via effective gauge fields. 
In-plane elastic strain alters the nearest-neighbor hopping amplitudes without breaking the planar embedding, generating pseudovector gauge potentials that act as pseudo-magnetic fields~\cite{Guinea2010,Levy2010}. 

Curved geometries can be treated using a covariant Dirac equation, in which the spin connection specifies parallel transport of the spinor~\cite{Cortijo2007,Iorio2014}. The role of Weyl symmetry in the massless description of curved graphene was examined by Iorio~\cite{Iorio2011}. The spin connection enters the kinetic term of the Hamiltonian without generating a sublattice mass. Changes in hopping amplitudes or sublattice on-site energies enter separately.

Graphene wormholes are topological defects where a curved bridge continuously connects two macroscopic layers~\cite{GonzalezHerrero2010}.
Nonhexagonal rings and reconstructed interfaces affect the lattice geometry and electronic states~\cite{Thrower1969,Stone1986,Yazyev2010}. Artificial honeycomb lattices offer a complementary setting for controlling Dirac quasiparticles~\cite{Gomes2012}.

The Hankel and hypergeometric scattering construction for hyperbolic wormholes is established in Refs.~\cite{pimsamarn2020,Naderi2024}, which use a common curvature scale on both sides and include an axial magnetic flux. Related work treats strain-dependent electronic states on a minimal-catenoid graphene wormhole~\cite{Silva2024}. We extend the hyperbolic scattering geometry to two independent curvature scales, with zero axial flux and elastic gauge fields neglected. We compare profiles with different asymmetries and pairs obtained by interchanging the throat ends. For each profile, we calculate the Landauer conductance and the local densities and currents of a specified incident channel. Asymmetric throats may be more realistic experimentally, since reproducing identical arrangements of nonhexagonal rings and other topological defects at both ends could be difficult.

The main contribution is a spatially resolved conductance response that distinguishes mirror throats with identical unperturbed conductance. We calculate the derivative with respect to a localized, azimuthally uniform scalar potential over energy and meridional position. At fixed displaced positions the mirror responses can have opposite signs, and the contrast survives a comparison at fixed throat length. The channel sum cancels the sublattice imbalance but retains this electrical response. We map the response over energy and position at fixed curvature-scale sum and test the sign contrast at a selected operating point with the throat length held fixed.

The remainder of this paper is organized as follows. Section~\ref{sec:model} introduces the model and covariant Dirac equation, and Secs.~\ref{sec:wavefunctions} and \ref{sec:tmatrix} develop the scattering eigenstates, transfer matrix and boundary conditions. Transport and transmission phase times are discussed in Sec.~\ref{sec:results}, followed by probability currents and sublattice observables in Sec.~\ref{sec:hydro}. Section~\ref{sec:gate} presents the spatially resolved electrical response and possible experimental probes. Consistency checks and the scope of the model are discussed in Sec.~\ref{sec:limits}, with the discussion and conclusions in Sec.~\ref{sec:conclusions}. Appendices~\ref{app:hypergeometric} and \ref{app:gate} give the hypergeometric derivation and the linear conductance-response equations.

\section{\label{sec:model}Model and Covariant Dirac Equation}

We consider two asymptotically flat graphene layers joined by a
structurally asymmetric wormhole described by the $(2{+}1)$-dimensional
line element
\begin{equation}
 ds^2 = -\vF^2dt^2 + du^2 + R^2(u)\,dv^2,
 \label{eq:metric}
\end{equation}
where $u$ is meridional arc length, $v$ is the azimuthal angle, and $t$ is physical time. We join two hyperbolic segments of the form used in Ref.~\cite{Naderi2024}, assigning them separate curvature scales:
\begin{equation}
 R(u) = \begin{cases}
  u - u_p + R_p, & u > u_p,\\[4pt]
  a\cosh(u/\rR), & 0 \le u \le u_p,\\[4pt]
  a\cosh(u/\rL), & u_m \le u < 0,\\[4pt]
  -(u-u_m)+R_m, & u < u_m,
 \end{cases}
 \label{eq:Ru}
\end{equation}
where $a$ is the fixed minimum radius of the throat. The parameters $r_{R}$ and $r_{L}$ define the curvature scales of the upper and lower throat segments. We quantify the geometric asymmetry through the ratio $\eta = \rR / \rL$. For $\eta = 1$, the throat profile is symmetric. The matching radii and connection points at the asymptotic flat layers are
\begin{align}
     u_p &= \rR\ln\!\Bigl(\sqrt{1+\rR^2/a^2} + \rR/a\Bigr), \quad R_p = \sqrt{a^2+\rR^2}, \nonumber \\
     u_m &= \rL\ln\!\Bigl(\sqrt{1+\rL^2/a^2} - \rL/a\Bigr), \quad R_m = \sqrt{a^2+\rL^2}.
     \label{eq:upm}
\end{align}
The functions $R$ and $R'$ are continuous at $u_m$, $0$, and $u_p$. The second derivative has finite jumps, but no Dirac delta functions occur in $R''$. In each curved segment the Gaussian curvature is $K=-R''/R=-1/r_j^2$, with $j=\mathrm{L},\mathrm{R}$; it vanishes in the flat layers. This metric is not a Euclidean minimal catenoid, even when $\eta=1$. In proper meridional coordinates, a minimal catenoid instead has $R(u)=\sqrt{a^2+u^2}$~\cite{Silva2024}.

\begin{figure*}[t]
 \centering
 \begin{minipage}[t]{0.57\textwidth}
  \vspace{0pt}
  {\sffamily\small (a) Surface Geometry\par}
  \vspace{4pt}
  \includegraphics[width=\linewidth]{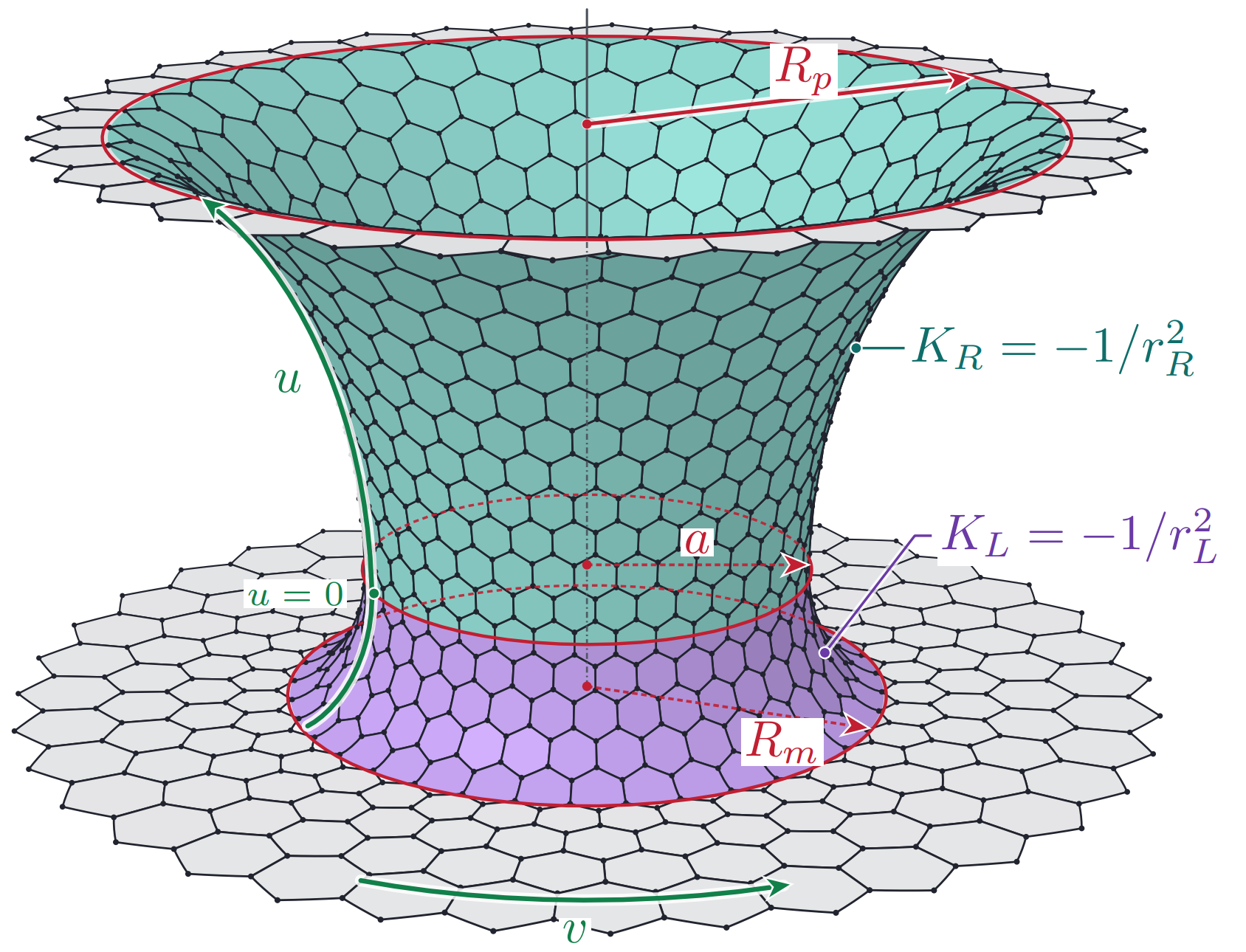}
 \end{minipage}\hfill
 \begin{minipage}[t]{0.41\textwidth}
  \vspace{0pt}
  {\sffamily\small (b) Meridional section\par}
  \vspace{36pt}
  % Numerically integrated meridional section, with the same metric parameters.
  \includegraphics[width=\linewidth]{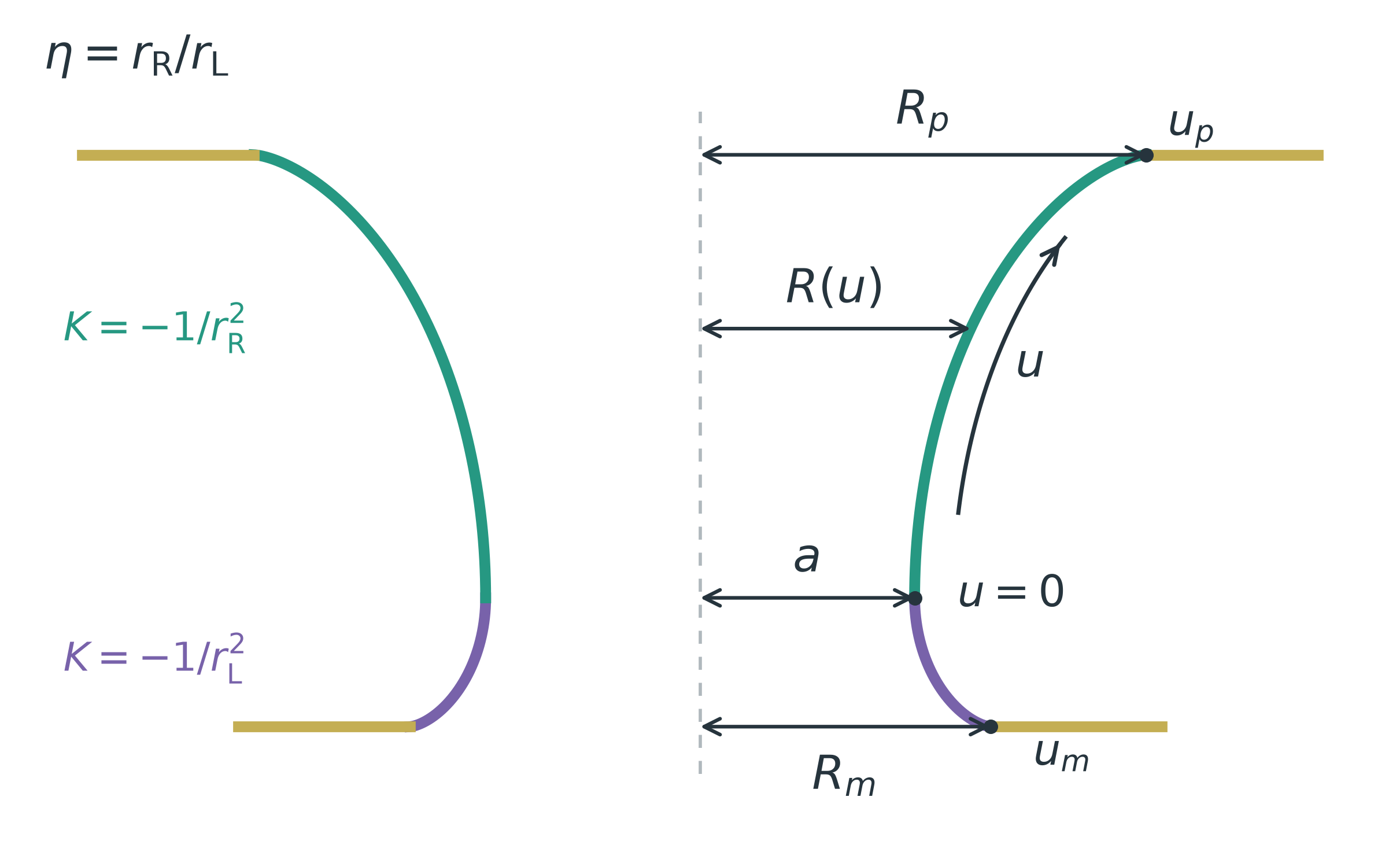}
 \end{minipage}
 \caption{Asymmetric throat with $r_{\mathrm R}>r_{\mathrm L}$. (a) Surface illustration identifying the minimum radius $a$, mouth radii $R_p$ and $R_m$, and the two curved segments. The flat layers are gray, the upper throat is teal, and the lower throat is purple. Green arrows indicate increasing meridional coordinate $u$ and periodic azimuth $v\in[0,2\pi)$, with $v\equiv v+2\pi$. The honeycomb overlay is schematic and does not specify an atomistically reconstructed lattice. (b) Meridional section obtained by numerically integrating the isometric embedding of Eqs.~\eqref{eq:metric}--\eqref{eq:upm}, with equal horizontal and vertical scales. Dimension arrows identify $a$, $R_p$, $R_m$, and $R(u)$; gold lines mark the flat layers. The $u$ arrow follows the meridian. Curvature labels in both panels refer to the intrinsic Gaussian curvature, $K_j=-R''/R=-1/r_j^2$ ($j=\mathrm R,\mathrm L$), so $r_j=1/\sqrt{-K_j}$ is an intrinsic curvature length, not a radial distance or the radius of curvature of the embedded meridional curve. Numerical scales are omitted.}
 \label{fig:schematic}
\end{figure*}

An isometric surface of revolution can be written as
\[
\begin{gathered}
\boldsymbol{x}(u,v)=(R(u)\cos v,R(u)\sin v,z(u)),\\
z'(u)=\sqrt{1-R'(u)^2}.
\end{gathered}
\]
The junctions in Eq.~\eqref{eq:upm} are precisely where $|R'|=1$, so this embedding is real throughout each throat segment and joins horizontal layers. Figure~\ref{fig:schematic} illustrates the asymmetric throat and identifies its geometric parameters and coordinates. Relating the continuum parameters $r_{\mathrm{L},\mathrm{R}}$ to an atomistic defect arrangement requires a microscopic model of the reconstructed lattice.

The low-energy dynamics of charge carriers on this manifold is governed
by the covariant massless Dirac
equation~\cite{Cortijo2007,Iorio2014}
\begin{equation}
 i\hbar\hat\gamma^a e^\mu_a\nabla_\mu\Psi=0.
\end{equation}
where $e_a^{\;\mu}$ are the inverse vielbeins, $\nabla_\mu = \partial_\mu
- \Gamma_\mu$ is the spinorial covariant derivative, and $\Gamma_\mu$
is the spin connection. For the unperturbed junction the external electromagnetic gauge field is set to zero ($A_\mu = 0$); the scalar perturbation introduced in Sec.~\ref{sec:gate} is treated separately. For the metric~\eqref{eq:metric} the only
non-vanishing spin connection is
\begin{equation}
 \Gamma_v = \tfrac{1}{2}\hat{\gamma}^1\hat{\gamma}^2 R'(u),
 \label{eq:spinconn}
\end{equation}
We choose $\hat\gamma^0=-i\sigma_3$, $\hat\gamma^1=\sigma_2$, and $\hat\gamma^2=-\alpha\sigma_1$, satisfying the Clifford algebra with signature $(-,+,+)$. The inverse vielbein is diagonal, with entries $(1/\vF,1,1/R)$. The index $\alpha=\pm1$ labels decoupled valley sectors, and the radial spinor is ordered as $\psi_\alpha=(\varphi_\alpha,\chi_{-\alpha})^T$, with $A$ and $B$ components in that order.

For the stationary state $\Psi=e^{-iEt/\hbar}e^{imv}\psi_\alpha(u)$, define $\mathcal A=\partial_u+R'/(2R)$ and $\mathcal B=\alpha m/R$. The coupled equations are
\[
 (\mathcal A-\mathcal B)\varphi_\alpha=ik\chi_{-\alpha},\qquad
 (\mathcal A+\mathcal B)\chi_{-\alpha}=ik\varphi_\alpha.
\]
Thus $[(\mathcal A+\mathcal B)(\mathcal A-\mathcal B)+k^2]\varphi_\alpha=0$. In particular, $-\mathcal B'=+\alpha mR'/R^2$, fixing the sign of the term linear in $m$ in

\begin{equation}
\begin{split}
 \biggl[\partial_u^2
  + \frac{R'}{R}\partial_u
  + k^2 + \frac{R''}{2R} - \frac{R'^2}{4R^2}\\
      + \frac{\alpha m R'}{R^2} - \frac{m^2}{R^2}
 \biggr]\varphi = 0,
\end{split}
 \label{eq:masterode}
\end{equation}
where $k = E/(\hbar v_{F})$ and $\alpha = \pm 1$ labels the decoupled valley sectors.

\section{\label{sec:wavefunctions}Scattering Eigenstates}

\subsection{Asymptotic flat regions}

In the flat layers $R'=\lambda=\pm1$ and $R''=0$, so
Eq.~\eqref{eq:masterode} reduces to the Bessel equation of order
\begin{equation}
 \nu_\lambda = \tfrac{1}{2}|\lambda - 2m\alpha|.
 \label{eq:nu}
\end{equation}
The asymptotic scattering states are expanded in the Hankel basis $H_\nu^{(1,2)}$, representing outgoing and incoming cylindrical waves. For an
electron injected from the upper layer ($\lambda=+1$), the physical
boundary conditions are an incident wave of unit amplitude plus a
reflected wave of amplitude $R_\alpha$ in the top layer, and a purely
transmitted wave of amplitude $T_\alpha$ in the bottom layer
\begin{align}
 \varphi_{\rm top} &= H_{\nu_+}^{(2)}(kR) + R_\alpha\,H_{\nu_+}^{(1)}(kR),
 \label{eq:phitop}\\
 \varphi_{\rm bot} &= T_\alpha\,H_{\nu_-}^{(1)}(kR).
 \label{eq:phibot}
\end{align}
The complementary spinor components $\chi$ are obtained by acting
with the first-order operator
$\chi_{-\alpha} = (ik)^{-1}[\lambda\partial_R +
(\lambda-2m\alpha)/(2R)]\varphi_\alpha$. Bessel recurrence relations reduce this to a single Hankel function of shifted order,
\begin{equation}
 \chi = -is_\alpha\, Z_{\nu - \lambda s_\alpha}(kR),
 \label{eq:chiflat}
\end{equation}
where $\nu=\nu_\lambda$ and $s_\alpha = \sgn(\lambda-2m\alpha)$ and $Z_\nu$ denotes either
$H_\nu^{(1)}$ or $H_\nu^{(2)}$ as appropriate. When $\lambda-2m\alpha=0$, we take $s_\alpha=+1$ in this recurrence; setting it to zero would incorrectly remove the lower component.

\subsection{Wormhole throat: Piecewise Gauss hypergeometric solutions}

In each curved segment, the complexified radial equation has regular singular points. Given the piecewise metric of Eq.~\eqref{eq:Ru}, we map Eq.~\eqref{eq:masterode} into the Gauss hypergeometric form independently for the $u>0$ and $u<0$ regions, using the continuations $X_{R,L}=\sinh(u/r_{R,L})$ and $z_{R,L}=\frac{1}{2}(1-iX_{R,L})$.

Factoring the local behavior at the regular singular points (see Appendix~\ref{app:hypergeometric}) yields the throat solutions $\varphi_{\rm th}(u) = \varphi_{\rm th}^{\rm R}(u)$ for $u \ge 0$ and $\varphi_{\rm th}^{\rm L}(u)$ for $u < 0$. The regional wavefunctions are
\begin{multline}
 \varphi_{\rm th}^{\rm R,L}(u) = A_{\rm in}^{\rm R,L}(1+iX_{\rm R,L})^{\gamma_1^{\rm R,L}}(1-iX_{\rm R,L})^{\gamma_2^{\rm R,L}} \\
 \times \,{}_2F_1\!\bigl(a_h^{\rm R,L},b_h^{\rm R,L};\,c_h^{\rm R,L};\,z_{\rm R,L}\bigr) \\
 + B_{\rm in}^{\rm R,L}(1+iX_{\rm R,L})^{\gamma_1^{\rm R,L}}(1-iX_{\rm R,L})^{\gamma_2^{\rm R,L}} \\
 \times \,z_{\rm R,L}^{1-c_h^{\rm R,L}}{}_2F_1\!\bigl(a_h^{\prime\rm R,L},b_h^{\prime\rm R,L};\,c_h^{\prime\rm R,L};\,z_{\rm R,L}\bigr),
 \label{eq:phithroat}
\end{multline}
where $\{A_{\rm in}^{\rm R,L}, B_{\rm in}^{\rm R,L}\}$ are the regional internal scattering amplitudes. The characteristic exponents describing the phase and decay near the regular singular points are $\gamma_1^{\rm R,L} = \tfrac{1}{4} + \tfrac{i\alpha m r_{\rm R,L}}{2a}$ and $\gamma_2^{\rm R,L} = \tfrac{1}{4} - \tfrac{i\alpha m r_{\rm R,L}}{2a}$
and the hypergeometric parameters are determined by the regional geometry and incident energy
$ a_h^{\rm R,L} = 1-ikr_{\rm R,L}$ and $
 b_h^{\rm R,L} = 1+ikr_{\rm R,L}$,
with $c_h^{\rm R,L} = 2\gamma_2^{\rm R,L}+1$, $a_h^{\prime\rm R,L}=a_h^{\rm R,L}-c_h^{\rm R,L}+1$, $b_h^{\prime\rm R,L}=b_h^{\rm R,L}-c_h^{\rm R,L}+1$, and $c_h^{\prime\rm R,L}=2-c_h^{\rm R,L}$~\cite{DLMF2024}.

\subsection{Linear independence and basis validity}

Both terms in Eq.~\eqref{eq:phithroat} share the same prefactor. For the two remaining Gauss solutions, linear independence for non-integer $c_h^{R,L}$ is ensured by the non-vanishing Wronskian
$W[\Phi_1,\Phi_2] \propto (1-c_h^{\rm R,L})\,z_{\rm R,L}^{-c_h^{\rm R,L}}(1-z_{\rm R,L})^{c_h^{\rm R,L}-a_h^{\rm R,L}-b_h^{\rm R,L}-1}$. In our case
$c_h^{\rm R,L} = 2\gamma_2^{\rm R,L}+1 = \tfrac{3}{2}-i\alpha m r_{\rm R,L}/a$, which has
non-zero imaginary part for $m\neq0$. Since we work with half-integer
modes $|m|\geq0.5$, $c_h^{\rm R,L}\notin\mathbb{R}$ and hence $c_h^{\rm R,L}\notin\mathbb{Z}$
for all parameter values used.

For $|z_{R,L}|>1$, we use the connection formula~\cite{DLMF2024}
\begin{equation}
\begin{split}
 {}_2F_1(a,b;c;z) &= \frac{\Gamma(c)\Gamma(b-a)}{\Gamma(b)\Gamma(c-a)} (-z)^{-a} \\
 &\quad \times {}_2F_1\!\left(a,a-c+1;a-b+1;\tfrac{1}{z}\right) \\
 &\quad + \frac{\Gamma(c)\Gamma(a-b)}{\Gamma(a)\Gamma(c-b)} (-z)^{-b} \\
 &\quad \times {}_2F_1\!\left(b,b-c+1;b-a+1;\tfrac{1}{z}\right)
\end{split}
 \label{eq:connection}
\end{equation}
to continue the solution analytically.

The $\chi$ spinor components in the throat follow from substituting
Eq.~\eqref{eq:phithroat} back into the first-order coupled Dirac
system locally for each region. Because $R = a\cosh(u/r_{\rm R,L})$ and $X_{\rm R,L} = \sinh(u/r_{\rm R,L})$, the longitudinal derivative transforms as $\partial_u = \frac{\sqrt{X_{\rm R,L}^2+1}}{r_{\rm R,L}}\partial_{X_{\rm R,L}}$, while $R'/(2R)=X_{\rm R,L}/[2r_{\rm R,L}\sqrt{X_{\rm R,L}^2+1}]$.
The resulting operator acting on $\varphi^{\rm R,L}_{\rm th}$ is
\begin{equation}
\begin{split}
\chi^{\rm R,L}_{\rm th} &= \frac{-i}{k r_{\rm R,L}\sqrt{X_{\rm R,L}^2+1}} \\
 &\quad \times \left[ \left(X_{\rm R,L}^2+1\right)\partial_{X_{\rm R,L}} + \frac{X_{\rm R,L}}{2} - \alpha \frac{m r_{\rm R,L}}{a} \right]\varphi^{\rm R,L}_{\rm th}
\end{split}
\label{eq:chithoat_op}
\end{equation}
The sign of the last term follows directly from $(\mathcal A-\mathcal B)\varphi=ik\chi$. We evaluate these analytic basis functions using \textsc{mpmath} at 60 decimal digits. Comparisons with 100-digit calculations and an independent integration of the first-order equations are described below.

\section{\label{sec:tmatrix}Transfer Matrix and Boundary Conditions}

\subsection{Angular boundary conditions and spinor definition}

We neglect intervalley mixing and work separately in each $\alpha$ sector. The angular spectrum depends on a boundary condition specified in addition to the metric. Throughout the calculations we choose the half-integer sector $m\in\mathbb Z+1/2$, corresponding to an antiperiodic angular envelope in the chosen rotating frame. In a flat layer, a single-valued envelope in a fixed Cartesian sublattice basis acquires this antiperiodicity when expressed in the local polar frame: the spinor rotation changes sign after $2\pi$~\cite{Ruegg2013}. Retaining that boundary condition across the junction gives $e^{2\pi i m}=-1$ without introducing an axial magnetic flux. Defect-induced holonomies can modify the angular spectrum~\cite{GonzalezHerrero2010,LammertCrespi2004,Vozmediano2010} and require microscopic information beyond the prescribed metric.

We use a continuum description away from atomic defects, where the geometric lengths exceed the lattice spacing. We neglect intervalley scattering at reconstructed junctions as an independent model assumption; its strength cannot be bounded from the separation between geometric and lattice scales alone. Strain-induced gauge fields and changes in hopping amplitudes are also omitted.

\subsection{T-matrix construction}

We impose continuity of the effective two-component Dirac spinor $(\varphi_\alpha,\chi_{-\alpha})^T$ at $u=u_p$, $u=0$, and $u=u_m$. This yields a sequence of $2\times2$ linear systems, which assemble into the block equations
\begin{equation}
\begin{split}
 \mathbf{M}_{\rm top} \begin{pmatrix}1\\R_\alpha\end{pmatrix} &= \mathbf{M}_{\rm th, R}(u_p) \begin{pmatrix}A_{\rm in}^{\rm R}\\B_{\rm in}^{\rm R}\end{pmatrix}, \\
 \mathbf{M}_{\rm th, R}(0) \begin{pmatrix}A_{\rm in}^{\rm R}\\B_{\rm in}^{\rm R}\end{pmatrix} &= \mathbf{M}_{\rm th, L}(0) \begin{pmatrix}A_{\rm in}^{\rm L}\\B_{\rm in}^{\rm L}\end{pmatrix}, \\
 \mathbf{M}_{\rm th, L}(u_m) \begin{pmatrix}A_{\rm in}^{\rm L}\\B_{\rm in}^{\rm L}\end{pmatrix} &= \mathbf{M}_{\rm bot} \begin{pmatrix}T_\alpha\\0\end{pmatrix}.
\end{split}
 \label{eq:matching}
\end{equation}
Using $\nu=\nu_+$ in the upper-layer matrix and $\nu=\nu_-$ in the lower-layer matrix, and suppressing the radial arguments, the
asymptotic matrices built from the Hankel functions and their chiral
partners [Eqs.~\eqref{eq:phitop}--\eqref{eq:chiflat}] are
\begin{equation}
 \mathbf{M}_{\rm top} =
 \begin{pmatrix}
  H_\nu^{(2)} & H_\nu^{(1)}\\[3pt]
  -is_\alpha H_{\nu-s_\alpha}^{(2)} & -is_\alpha H_{\nu-s_\alpha}^{(1)}
 \end{pmatrix},
 \label{eq:Mtop}
\end{equation}
\begin{equation}
 \mathbf{M}_{\rm bot} =
 \begin{pmatrix}
  \tilde H_\nu^{(1)} & \tilde H_\nu^{(2)}\\[3pt]
  -is_\alpha' \tilde H_{\nu+s_\alpha'}^{(1)} &
  -is_\alpha' \tilde H_{\nu+s_\alpha'}^{(2)}
 \end{pmatrix},
 \label{eq:Mbot}
\end{equation}
where tildes denote evaluation at $k\Rm$ and
$s_\alpha' = \sgn(-1-2m\alpha)$ for the lower layer, with $s_\alpha'=+1$ when $-1-2m\alpha=0$, as in Eq.~\eqref{eq:chiflat}.
We invert $\mathbf{M}_{\rm top}$ analytically using the Wronskian identity $W[H_{\nu}^{(1)},H_{\nu}^{(2)}]=-4i/(\pi kR_p)$. The regional throat matrices $\mathbf{M}_{\rm th, R}(u_p)$ and $\mathbf{M}_{\rm th, L}(u_m)$ are constructed by evaluating Eqs.~\eqref{eq:phithroat} and \eqref{eq:chithoat_op} at $X_{\rm R}(u_p) = \sinh(u_p/\rR)$ and $X_{\rm L}(u_m) = \sinh(u_m/\rL)$. Eliminating the internal amplitudes $A_{\rm in}^{\rm R,L}$ and $B_{\rm in}^{\rm R,L}$ yields the global T-matrix
\begin{equation}
 \begin{split}
 \mathbf{W} &= \mathbf{M}_{\rm top}^{-1} \cdot \mathbf{M}_{\rm th, R}(u_p) \cdot \left[ \mathbf{M}_{\rm th, R}^{-1}(0) \cdot \mathbf{M}_{\rm th, L}(0) \right] \\
 &\quad \cdot \mathbf{M}_{\rm th, L}^{-1}(u_m) \cdot \mathbf{M}_{\rm bot}.
\end{split}
 \label{eq:Tmatrix}
\end{equation}
The unit-coefficient Hankel spinors on either layer carry radial fluxes of equal magnitude, $4\vF/(\pi k)$ per unit azimuthal angle. Hence $T_\alpha=1/W_{11}$ and $R_\alpha=W_{21}/W_{11}$ give the flux probabilities $|T_\alpha|^2$ and $|R_\alpha|^2$. The mapped coordinate obeys $\operatorname{Re}z=1/2$ for real $u$, so the matching contour avoids the principal Gauss branch cut $[1,\infty)$.

\emph{Formal Hermiticity and flux conservation.}---The effective
first-order radial Dirac operator on the wormhole is
$\hat{D} = \sigma^1\bigl(\partial_u + R'/(2R)\bigr) +i
(\alpha\sigma^2 m)/R$.
Under the $u$-line measure weighted by $R(u)$ (the geometric inner
product induced by the metric~\eqref{eq:metric}),
$\langle\Psi_1,\Psi_2\rangle = \int R(u)\,\Psi_1^\dagger\Psi_2\,du$,
the operator $\hat{D}$ is formally \emph{skew-Hermitian}. A direct
integration by parts of the $\sigma^1\partial_u$ term gives
\begin{equation}
\begin{split}
 &\int R\,\Psi_1^\dagger\sigma^1\mathcal A\Psi_2\,du
 = \bigl[R\,\Psi_1^\dagger\sigma^1\Psi_2\bigr]_{\rm bdry} \\
 &\qquad - \int R\,(\mathcal A\Psi_1)^\dagger \sigma^1\Psi_2\,du,
\end{split}
 \label{eq:ibp}
\end{equation}
where we used $(\sigma^1)^\dagger=\sigma^1$ and
$\partial_u(R\Psi_1^\dagger\sigma^1\Psi_2)=R'(\Psi_1^\dagger\sigma^1\Psi_2)
+R[(\partial_u\Psi_1)^\dagger\sigma^1\Psi_2+\Psi_1^\dagger\sigma^1(\partial_u\Psi_2)]$.
The $i\alpha\sigma^2 m/R$ term is intrinsically skew-Hermitian. Hence the Hamiltonian $-i\hbarvF\hat{D}$ is formally Hermitian under $\langle\,\cdot\,,\,\cdot\,\rangle$.

For spinors continuous at each interface, the internal boundary terms in Eq.~\eqref{eq:ibp} cancel because $R$ is also continuous. Formal Hermiticity applies, for example, to smooth compactly supported states with these matching conditions. Scattering eigenstates are not square integrable, so their flux balance is more directly obtained from the coupled equations: $\partial_u(R\psi^\dagger\sigma^1\psi)=0$.

For incidence from the upper layer the net flux is negative in the $u$ direction. In the normalization of Eqs.~\eqref{eq:phitop} and \eqref{eq:phibot},
\[
 \mathcal F=R J^u=-\frac{4\vF}{\pi k}(1-|R_\alpha|^2)
 =-\frac{4\vF}{\pi k}|T_\alpha|^2.
\]
This establishes $|R_\alpha|^2+|T_\alpha|^2=1$ without assuming the vanishing of scattering-state boundary terms at infinity.

\section{\label{sec:results}Transport and Transmission Phase Time}

\subsection{Transmission probability and angular momentum filtering}

Figure~\ref{fig:transport} compares three allowed channels, $m=0.5,1.5,2.5$, for symmetric profiles. We use $\hbarvF=0.658$~eV\,nm and $\alpha=1$. Unless varied, $E=120$~meV, $a=5$~nm, and $r_{\mathrm L}=r_{\mathrm R}=r=3$~nm. The largest plotted wave number is $k\simeq0.46$~nm$^{-1}$, below the inverse lattice spacing; the calculation assumes the linear Dirac dispersion on this range.

The suppression of higher-$|m|$ channels can be understood by writing $f=\sqrt R\,\varphi$. Its equation is $f''+[k^2-V_\varphi(u)]f=0$, with
\[
 V_\varphi(u)=\frac{m^2}{R^2}-\frac{\alpha m R'}{R^2}.
\]
The characteristic angular energy at the waist is $E_m=\hbarvF|m|/a$. At $a=5$~nm it is approximately 66~meV for $m=0.5$ and 329~meV for $m=2.5$. Their use as channel thresholds is only approximate, since transmission also depends on the derivative term and finite barrier extent. All fixed-$m$ channels are propagating sufficiently far into either flat layer; a suppressed channel is therefore not evanescent throughout the manifold.

The plotted energy dependence shows the rise from suppressed to high transmission, while the length dependence distinguishes the weakly reflected lowest channel from more strongly attenuated higher channels.

\subsection{Transmission phase time}

We calculate the derivative of the transmission phase in the Hankel basis of Eqs.~\eqref{eq:phitop} and \eqref{eq:phibot}~\cite{Wigner1955}:
\begin{equation}
 \tau_W=\hbar\,\operatorname{Im}\!\left(\frac{1}{T_\alpha}
 \frac{dT_\alpha}{dE}\right).
 \label{eq:wigner}
\end{equation}
For example, at $E=120$~meV, $a=5$~nm, and $r=3$~nm, the $m=0.5$ channel has $|T|^2=0.9502$ and $\tau_W=-6.307$~fs, whereas $m=2.5$ has $|T|^2=1.44\times10^{-4}$ and $\tau_W=0.049$~fs. The negative phase derivative occurs here in the high-transmission channel; its sign is insufficient to identify a tunneling regime. We approximate $dT_\alpha/dE$ by a centered finite difference with step $\Delta E=10^{-5}$~eV and use $\hbar=0.6582119569$~eV\,fs to express $\tau_W$ in femtoseconds. Representative step-size and precision checks are reported in Sec.~\ref{sec:limits}.

\begin{figure*}[t]
 \centering
 \includegraphics[width=\textwidth]{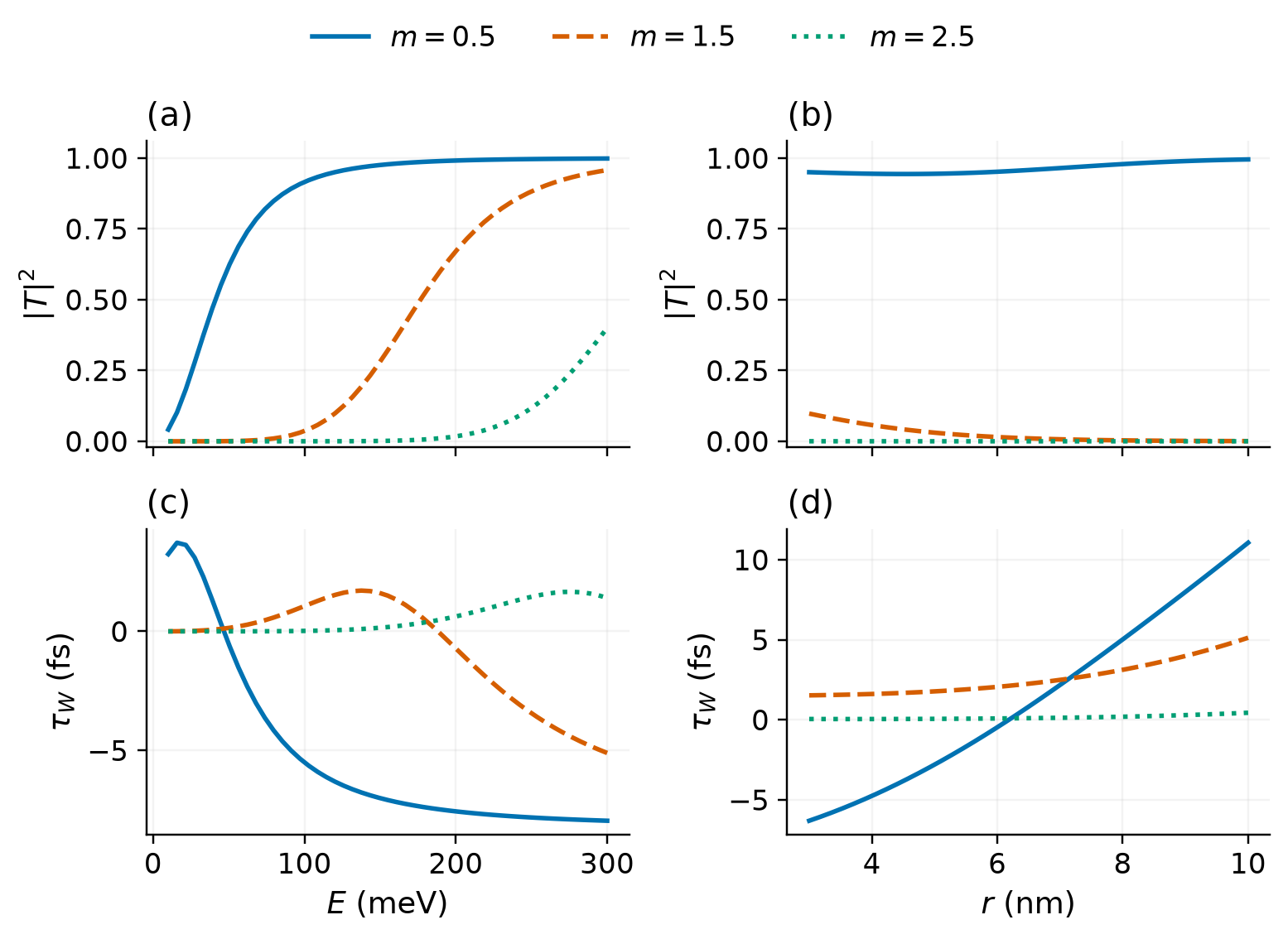}
 \caption{Symmetric-throat transport for $m=0.5,1.5,2.5$ and $\alpha=1$. (a,b) Transmission probability and (c,d) transmission phase time in the specified Hankel convention. The left column varies energy at $a=5$~nm and $r=3$~nm; the right column varies $r$ at $a=5$~nm and $E=120$~meV.}
 \label{fig:transport}
\end{figure*}

The asymptotic phase convention enters $\tau_W$ through the transmission amplitude: an energy-dependent rephasing contributes an additional phase derivative. To obtain an excess travel time for specified source and detector positions, one must include the corresponding reference propagation phase. Figures~\ref{fig:transport}(c,d) and \ref{fig:global_trans}(b) give the scattering phase derivatives in the chosen convention.

We compare asymmetric profiles at fixed $a$ and $r_{\rm eff}=r_{\mathrm L}+r_{\mathrm R}=13.68$~nm in Fig.~\ref{fig:global_trans}. Interchanging $r_{\mathrm L}$ and $r_{\mathrm R}$ sends $\eta$ to $1/\eta$. The simultaneous transformation $u\mapsto-u$ and $\alpha\mapsto-\alpha$ leaves Eq.~\eqref{eq:masterode} invariant after exchanging the two ends. Together with equal left-to-right and right-to-left transmission probabilities for the lossless two-channel scattering matrix, this gives
\[
 |T_m(\eta,\alpha)|^2=|T_m(1/\eta,-\alpha)|^2.
\]
The channel pairing $H_{\alpha,-m}=\sigma_1H_{\alpha,m}\sigma_1$ derived in Sec.~\ref{sec:hydro}, together with the dependence on $\alpha m$, gives equal transmission probabilities in opposite valley sectors at fixed $m$. Combining it with the mirror relation explains the coincidence at fixed $\alpha$ in Fig.~\ref{fig:global_trans}(a). In the Hankel phase convention used here, the mirror partners also have equal complex transmission amplitudes, $T_m(\eta,\alpha)=T_m(1/\eta,\alpha)$, and hence equal transmission phase times. For two solutions $\boldsymbol f$ and $\boldsymbol g$ of the rescaled system~\eqref{eq:gate_radial}, $W[\boldsymbol f,\boldsymbol g]=f_1g_2-f_2g_1$ is conserved because $\operatorname{tr}\mathsf A_m=0$. The rescaled Hankel basis matrices in Eqs.~\eqref{eq:Mtop}--\eqref{eq:Mbot} both have determinant $4/(\pi k)$, giving equal upper- and lower-incident transmission amplitudes. The mirror map $\boldsymbol f(u)\mapsto\sigma_2\boldsymbol f(-u)$ interchanges these incidences and takes $\eta$ into $1/\eta$ while preserving the transmission coefficient. It also applies to a scalar gate when its profile is reflected. The mirror partners can have different reflection phases. These relations apply when the two curvature scales are interchanged at fixed $a$; arbitrary changes in $\eta$ can change the transmission.

\subsection{Macroscopic conductance}

To evaluate transport in a multi-mode injection regime, we compute the zero-temperature macroscopic conductance using the Landauer-B\"uttiker formalism~\cite{Datta1995,Landauer1957,Buttiker1986}:
\begin{equation}
 G(E) = G_0 \sum_{m} |T_m(E)|^2,
 \label{eq:landauer}
\end{equation}
where $G_0=4e^2/h$ includes spin and valley degeneracy. We calculate one valley ($\alpha=1$) and sum positive and negative half-integer $m$. Since the radial equations depend on $\alpha m$, changing valley permutes the terms of this complete angular sum. The factor of four therefore does not count the two sublattice components as separate channels.

\begin{figure*}[t]
 \centering
 \includegraphics[width=\textwidth]{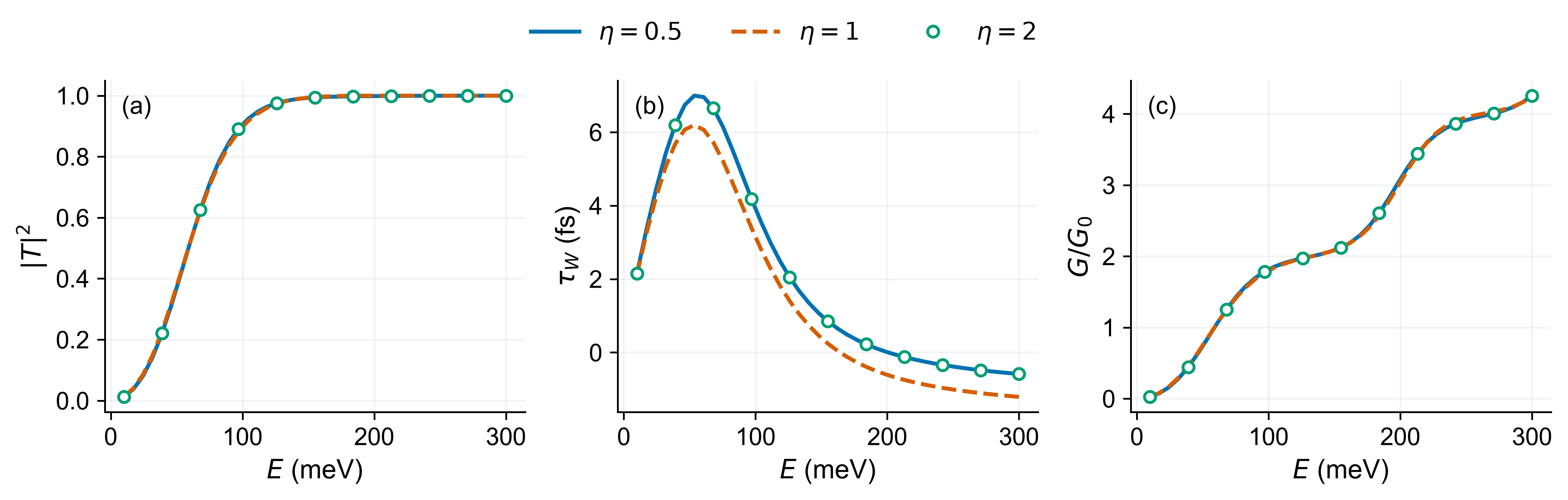}
 \caption{Dependence on asymmetry at fixed $a=5$~nm and $r_{\rm eff}=13.68$~nm. (a) Transmission and (b) transmission phase time for $m=0.5$, $\alpha=1$. (c) Landauer conductance including both signs of the half-integer angular modes. The $\eta=0.5$ and $2$ profiles are mirror partners; the $\eta=0.5$ results are shown by a blue solid line and the coincident $\eta=2$ results by green open circles at selected calculated points. The symmetric profile is shown by an orange dashed line.}
 \label{fig:global_trans}
\end{figure*}

The sum is evaluated until three successive pairs $(m,-m)$ contribute less than $10^{-7}$ each; a tighter-cutoff comparison is given below. Higher angular channels contribute progressively as their transmission increases. The resulting staircase-like behavior of $G(E)$ in Fig.~\ref{fig:global_trans}(c) reflects the successive opening of angular transport channels as the Fermi energy crosses their characteristic scales. This is analogous to conductance quantization in quantum point contacts~\cite{vanWees1988}, with the half-integer angular modes $m$ playing the role of transverse channels. It also resembles the edge-channel picture of quantum Hall transport~\cite{Buttiker1988QH}, although here the structure is purely geometric and involves neither Landau levels nor a magnetic field. Using the mirror relation and relabeling $m\mapsto-m$ gives $G(\eta)=G(1/\eta)$ at fixed $a$ and $r_{\rm eff}$. The symmetric profile need not have the same conductance as an asymmetric pair. Figure~\ref{fig:global_trans}(c) compares these cases. At $E=150$~meV, the mirror pair has $G/G_0=2.08457$, compared with $2.08609$ for the symmetric profile. Thus the conductance differs by about $0.073\%$ in this example, whereas the phase times for $m=0.5$ are approximately $1.003$ and $0.394$~fs, respectively.

\section{\label{sec:hydro}Probability Currents and Sublattice Observables}

\subsection{Density, current, and projection}

For a specified incident channel we define $\rho=\psi^\dagger\psi$. The probability in a surface element is proportional to $\rho R\,du\,dv$. The current components in the representation introduced above are
\begin{align}
 J^u&=\vF\,\psi^\dagger\sigma_1\psi,
 \label{eq:Ju}\\
 J^v&=\frac{\alpha\vF}{R}\,\psi^\dagger\sigma_2\psi.
 \label{eq:Jv}
\end{align}
They obey $\partial_t\rho+R^{-1}\partial_u(RJ^u)+\partial_vJ^v=0$. The physical azimuthal component in an orthonormal frame is $J^{\hat v}=RJ^v$; it must be distinguished from the coordinate component $J^v$.

The local plots use a unit incoming Hankel coefficient at the same energy, $E=150$~meV, for all geometries. The corresponding incident flux has the common magnitude $4\vF/(\pi k)$. The densities use this incident-amplitude normalization; the scattering states have no unit-total-probability normalization.

\begin{figure*}[t]
 \centering
 \includegraphics[width=\textwidth]{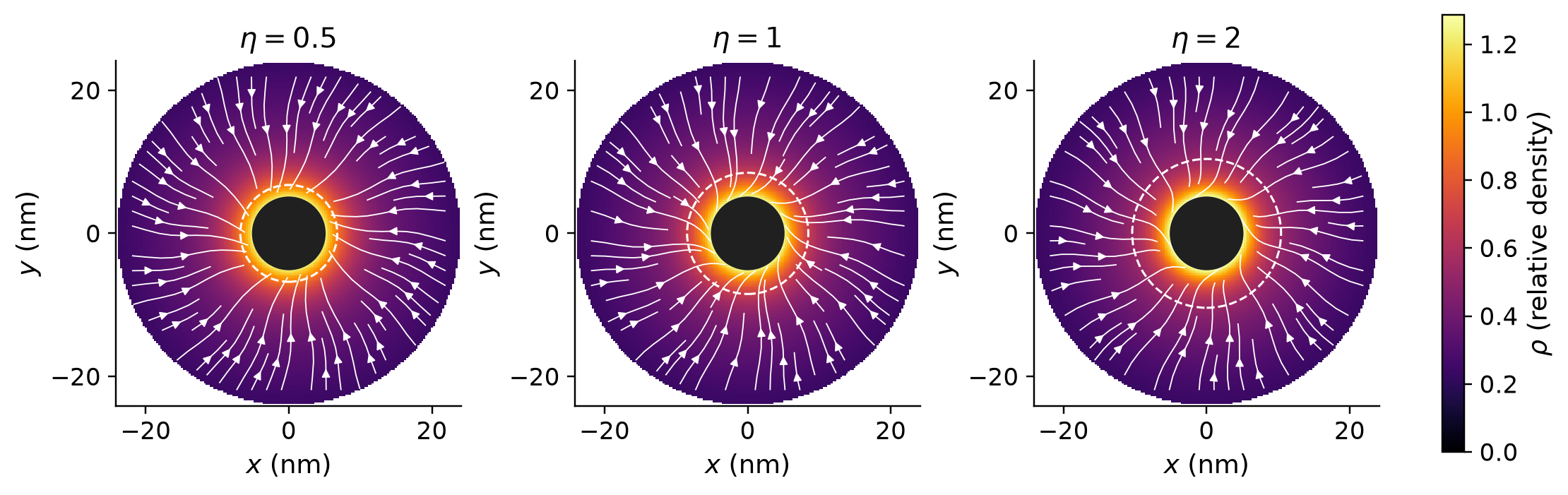}
 \caption{Upper-half probability density and projected current for $E=150$~meV, $a=5$~nm, $r_{\rm eff}=13.68$~nm, $m=0.5$, and $\alpha=1$. All panels use the same incident-amplitude normalization and color scale. Arrows indicate the projected current direction. The dark disk excludes $R<a$, and dashed circles mark the mouths $R=R_p$.}
 \label{fig:vortex}
\end{figure*}

Figure~\ref{fig:vortex} shows the upper half ($u\geq0$), where projection to the $(x,y)$ plane is single valued. The projected radial and azimuthal current components are $(R'J^u,RJ^v)$. In Cartesian coordinates,
\[
 \boldsymbol J_{\rm proj}=R'J^u\boldsymbol e_R+RJ^v\boldsymbol e_v.
\]
The figure uses the exact piecewise derivative, including $R'=1$ in the upper flat layer, and a common density color scale. For upper-layer incidence the net meridional flow is toward decreasing $u$, so the projected streamlines turn inward toward the throat. Their azimuthal component reflects the angular structure of the selected channel. At $u=0$, the projected radial component vanishes because $R'=0$; the nonzero conserved flux continues through the throat along the surface. The surface current remains finite through the waist.

\subsection{Spin connection and sublattice observables}

The contribution of $\Gamma_v$, proportional to $\sigma_3$ in this representation, enters the Hamiltonian through its contraction with the Dirac matrices. The contraction in the covariant equation gives
\[
 -\frac{\hat\gamma^2}{R}\Gamma_v
 =-\frac{R'}{2R}\hat\gamma^2\hat\gamma^1\hat\gamma^2
 =\frac{R'}{2R}\hat\gamma^1.
\]
Converting the time-dependent equation to Hamiltonian form therefore yields
\begin{equation}
 H_\alpha=-i\hbarvF\sigma_1\left(\partial_u+\frac{R'}{2R}\right)
       +\hbarvF\frac{\alpha m}{R}\sigma_2.
 \label{eq:Hgeom}
\end{equation}
The radial kinetic term contains the connection required for Hermiticity with the geometric measure (Sec.~\ref{sec:tmatrix}), with no additional real $\sigma_3 R'/R$ potential; the same kinetic structure appears in covariant treatments of catenoid graphene~\cite{Silva2024}.

The transformation $\widetilde\psi=\sqrt R\,\psi$ maps the radial inner product to the flat measure and gives
\[
 \widetilde H_\alpha=-i\hbarvF\sigma_1\partial_u
          +\hbarvF\frac{\alpha m}{R}\sigma_2.
\]
Integration of $R'/(2R)$ gives the amplitude factor in this rescaling. No real dynamical phase from a sublattice mass is generated. Geometry enters the rescaled states through the radius-dependent angular term and matching conditions, and the physical surface densities through the radial rescaling factor.

In the fixed component order $(A,B)$, the sublattice density difference is
\begin{equation}
 s_z(u)=\psi^\dagger\sigma_3\psi=|\varphi|^2-|\chi|^2.
 \label{eq:Pz}
\end{equation}
The locally normalized polarization is instead $p_z=s_z/\rho$, with $-1\leq p_z\leq1$ wherever $\rho>0$. Positive values denote a larger $A$-component weight. Angular dependence and interference in a massless scattering state can produce unequal local component weights and hence a nonzero sublattice density difference.

\begin{figure*}[t]
 \centering
 \includegraphics[width=\textwidth]{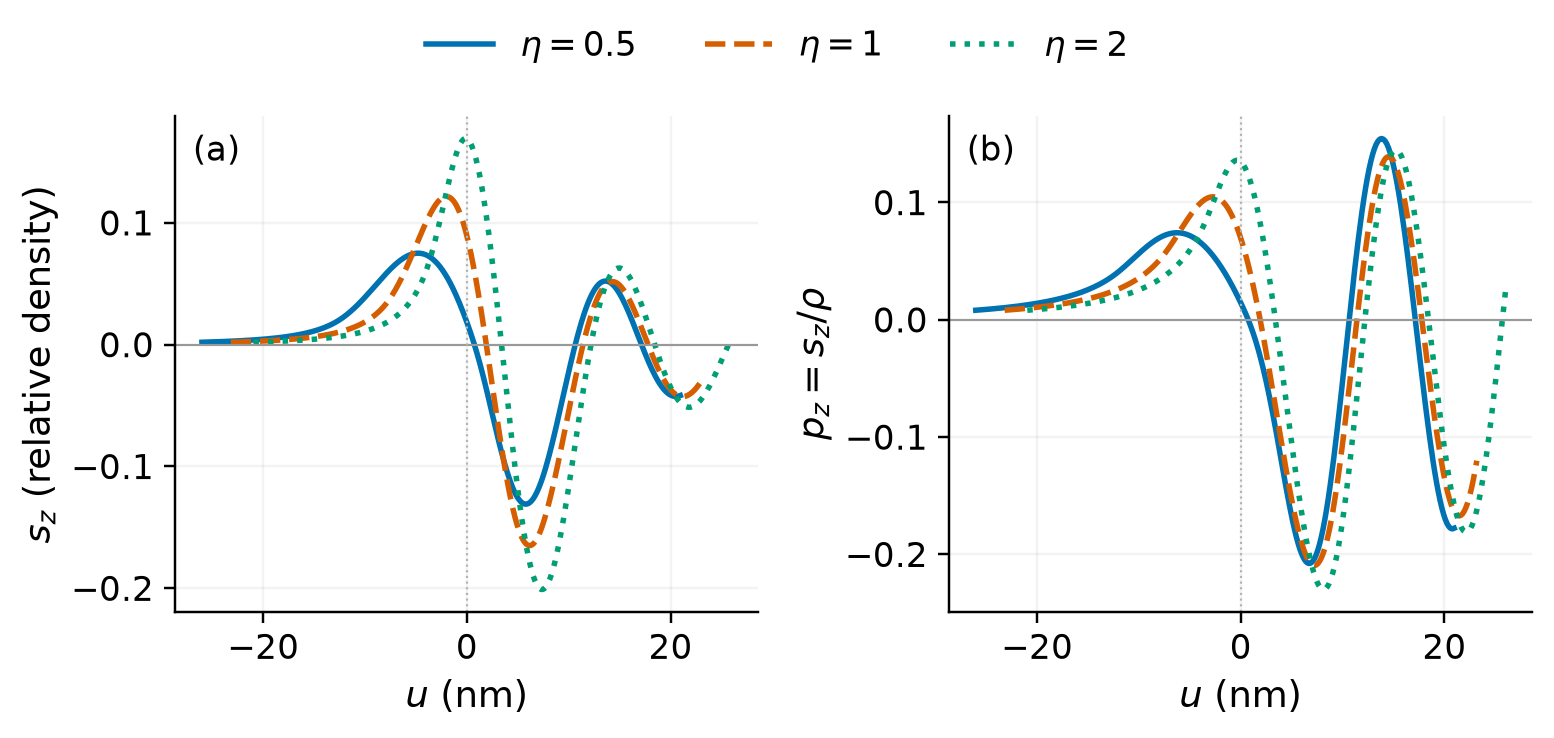}
 \caption{(a) Sublattice density difference $s_z=|\varphi|^2-|\chi|^2$ and (b) normalized polarization $p_z=s_z/\rho$, for the same incident channels and normalization as Fig.~\ref{fig:vortex}. The profiles include both flat layers and the throat; $u=0$ is the waist. Positive values correspond to the $A$ sublattice.}
 \label{fig:pseudospin}
\end{figure*}

Figure~\ref{fig:pseudospin} compares both quantities for upper-layer incidence. On the sampled profiles, the most negative $s_z$ values are approximately $-0.131$, $-0.165$, and $-0.201$ for $\eta=0.5$, $1$, and $2$, respectively. These extrema indicate a larger local $B$-component density and depend on the stated incident-amplitude normalization. The $p_z$ curves have a different relative variation because the normalization includes the local total density. The curves depend on the geometry and on the boundary condition selecting the incoming state. A geometric reflection exchanges the two scattering ends; it does not require an odd $s_z(u)$ for a fixed one-sided incident state, even when $\eta=1$. The density difference decays with cylindrical spreading in the far field, whereas its ratio to $\rho$ can retain incident--reflected interference oscillations.

Conjugation of Eq.~\eqref{eq:Hgeom} by the first Pauli matrix reverses the angular label. Writing its angular label explicitly,
\[
 H_{\alpha,-m}=\sigma_1 H_{\alpha,m}\sigma_1.
\]
For the same incidence direction and flux normalization, the paired radial scattering spinors are related by $\sigma_1$ up to an overall phase. Consequently,
\[
 \begin{gathered}
 \rho_{\alpha,-m}=\rho_{\alpha,m},\qquad
 s_{z,\alpha,-m}=-s_{z,\alpha,m},\\
 J^{\hat v}_{\alpha,-m}=-J^{\hat v}_{\alpha,m}.
 \end{gathered}
\]
The radial current is unchanged. Equal incoherent occupation of the $m$ and $-m$ channels therefore cancels both the net sublattice polarization and the azimuthal circulation, even for an asymmetric throat. A finite summed sublattice contrast requires either unequal sector weights or a perturbation that breaks the pairing symmetry.

\section{\label{sec:gate}Spatially Resolved Electrical Response}

Mirror junctions with equal unperturbed conductance can have different local scattering profiles. Equal occupation of opposite angular channels cancels the net sublattice polarization. We apply the same scalar ring gate at a selected position in each mirror junction and calculate the conductance change as the gate moves along the surface. Conductance responses to weak electrostatic probes are established in scanning gate theory, including graphene nanostructures~\cite{Gorini2013,Chen2023}.

We supplement Eq.~\eqref{eq:Hgeom} with a prescribed scalar potential,
\begin{equation}
\begin{split}
 H_{\alpha,m}^{(U)}&=H_{\alpha,m}+U_0g(u;u_0,w)\,\mathbb I,\\
 g(u;u_0,w)&=\exp\!\left[-\frac{(u-u_0)^2}{2w^2}\right].
\end{split}
\label{eq:gate_potential}
\end{equation}
The potential is localized in meridional arc length and uniform in $v$, so it represents an idealized ring gate and preserves the separated angular channels. $U_0$ is an energy, $w$ is its meridional width, and $u_0$ locates its center on the surface. We compare geometries at the same effective potential amplitude $U_0$ and width $w$. The external charge needed to realize that profile can differ between geometries. The relation $H_{\alpha,-m}^{(U)}=\sigma_1H_{\alpha,m}^{(U)}\sigma_1$ remains valid: the paired sublattice imbalances continue to cancel under equal occupation while the conductance responds to the gate.

With $G_{\rm empty}(E)=G(E,U_0=0)$ and $\Delta G=G-G_{\rm empty}$, we define
\begin{equation}
\begin{split}
 \chi_G(E,u_0)&=\left.\frac{\partial [G(E,U_0,u_0)/G_0]}{\partial U_0}\right|_{0},\\
 \frac{G-G_{\rm empty}}{G_0}&=U_0\chi_G(E,u_0)+O(U_0^2).
\end{split}
\label{eq:gate_response}
\end{equation}
The finite-potential curves express $\Delta G$ in units of $G_0=4e^2/h$; percentage changes quoted below are relative to $G_{\rm empty}$. The derivative is obtained from the first-order Dirac equation with the same Hankel boundary convention and angular sum as the unperturbed calculation. Appendix~\ref{app:gate} derives the variational equations and describes their numerical checks. The response is determined by the potential-induced changes of the scattering amplitudes~\cite{Gorini2013}.

\begin{figure*}[t]
 \centering
 \includegraphics[width=\textwidth]{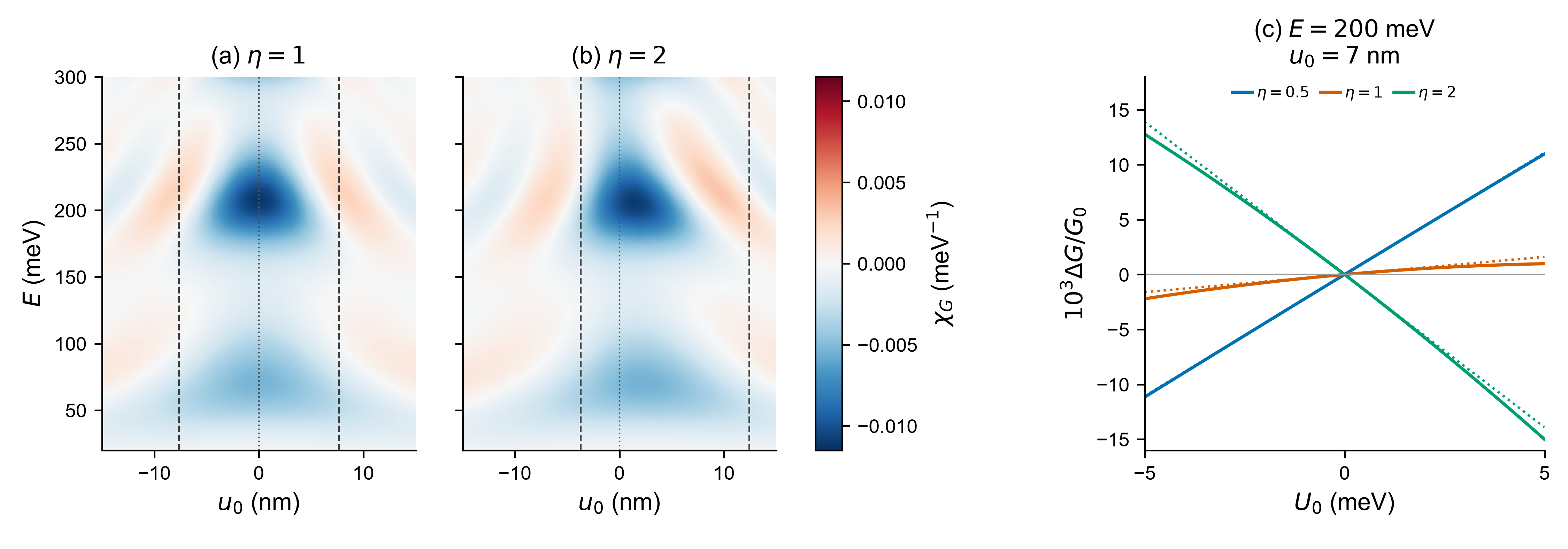}
 \caption{Conductance response to the scalar ring gate of Eq.~\eqref{eq:gate_potential}, for $a=5$~nm, $r_{\rm eff}=13.68$~nm and $w=1$~nm. (a,b) $\chi_G(E,u_0)$ for the symmetric and asymmetric profiles, on the same color scale. Blue and red indicate negative and positive response to increasing $U_0$. Dashed lines mark the mouths; the dotted line marks the waist. Each map uses $281\times301$ directly calculated points. The $\eta=0.5$ map follows by reflecting panel (b) about $u_0=0$. (c) Conductance change at $E=200$~meV and $u_0=7$~nm: solid curves are full finite-potential solutions and dotted lines show Eq.~\eqref{eq:gate_response}. The gate center is in a flat region for $\eta=0.5$ and a curved region for $\eta=1,2$. The scan position is meridional arc length on the surface, not the axial position of a molecule.}
 \label{fig:gate_response}
\end{figure*}

Figure~\ref{fig:gate_response} shows the response for $a=5$~nm, $r_{\rm eff}=13.68$~nm and $w=1$~nm. We sample $20\leq E\leq300$~meV and $-15\leq u_0\leq15$~nm with steps of $1$~meV and $0.1$~nm, respectively. The broad response bands follow the energy dependence of the angular channels: at the waist of the $\eta=2$ geometry, the $m=\pm0.5$ pair dominates near $60$~meV, while $m=\pm1.5$ contributes about $98\%$ of $\chi_G$ at $200$~meV. The bands occur on the angular-barrier scales discussed in Sec.~\ref{sec:results}.

Reflecting the geometry and the potential together gives
\begin{equation}
 \chi_{G,\eta}(E,u_0)=\chi_{G,1/\eta}(E,-u_0).
 \label{eq:gate_mirror}
\end{equation}
Thus the response of the symmetric junction is even in $u_0$, and mirror partners agree for a gate centered at the waist. At a fixed displaced position their responses need not coincide. For example, at $E=200$~meV and $u_0=7$~nm, the complete finite-potential solution with $U_0=+1$~meV gives $\Delta G/G_{\rm empty}=+0.0718\%$ for $\eta=0.5$ and $-0.0918\%$ for $\eta=2$, although both have $G_{\rm empty}/G_0=3.08066$. The symmetric response is $+0.00976\%$. By Eq.~\eqref{eq:gate_mirror}, comparing mirror throats at the same $u_0$ is equivalent to comparing the response of one asymmetric throat at $u_0$ and $-u_0$. The gate supplies the spatial reference. Two-terminal reciprocity concerns interchange of the contacts at a fixed potential profile.

In this example, $u_0=7$~nm lies in the flat layer for $\eta=0.5$ and in the curved segment for $\eta=2$. A weaker sign contrast also occurs near the edge of the common throat interval, $|u_0|\lesssim3.73$~nm: at $E=252$~meV and $u_0=3.7$~nm, $\chi_{G,0.5}=+1.3225\times10^{-4}$~meV$^{-1}$ and $\chi_{G,2}=-5.2793\times10^{-4}$~meV$^{-1}$. Both gate centers are then inside the throats, but the finite-width Gaussian extends into the flat layer of the shorter segment. Within the sampled energy range, no opposite-sign response occurs for $|u_0|\leq3$~nm at $w=1$~nm, although the magnitudes differ. At $E=200$~meV and $u_0=2$~nm, the responses are $-6.9190\times10^{-3}$ and $-1.0673\times10^{-2}$~meV$^{-1}$ for $\eta=0.5$ and $2$, respectively. The sign contrast near the mouths does not require equal gate amplitudes in the linear regime, provided both perturbations have the same known sign and their spatial profiles are maintained; magnitude comparisons require amplitude calibration.

For the fixed-length comparison we keep $a$ and the meridional throat length
\begin{equation}
 L=u_p-u_m=\rL\operatorname{arsinh}(\rL/a)
   +\rR\operatorname{arsinh}(\rR/a)
 \label{eq:gate_length}
\end{equation}
fixed at the symmetric-reference value $15.3112$~nm. For $\eta=2$, this gives $(\rL,\rR)=(4.41816,8.83633)$~nm. At $E=200$~meV, $u_0=7$~nm and $w=1$~nm, the mirror responses remain opposite in sign: $\chi_{G,0.5}=+2.2403\times10^{-3}$~meV$^{-1}$ and $\chi_{G,2}=-2.5188\times10^{-3}$~meV$^{-1}$. The sign difference at this operating point also persists for $w=0.5$ and $2$~nm in the fixed-$r_{\rm eff}$ family. The response magnitude varies with energy, gate position and width. At fixed $r_{\rm eff}$, asymmetry shifts the position and energy of the strongest response, while the maximum $|\chi_G|$ is about $1.5\%$ smaller for the asymmetric pair than for the symmetric profile in the sampled energy--position domain.

\subsection{Possible experimental probes}

Sweeping the Fermi energy in a two-terminal measurement between the layers probes the successive angular-channel contributions to conductance in Fig.~\ref{fig:global_trans}(c). A test of $G(\eta)=G(1/\eta)$ compares junctions with interchanged curvature scales at fixed $a$, under comparable contact coupling and angular boundary conditions. A controlled local gate would add the spatially selective comparison in Fig.~\ref{fig:gate_response}; its potential on the curved surface would need to be calibrated to relate a gate voltage to $U_0$ and $w$. The comparison with the ideal Landauer conductance requires separating the resistance of the layer contacts and disorder contributions from transport through the throat.

Tunneling spectroscopy near the mouths probes the local density of states, obtained from the full spectral sum described below. Sublattice-sensitive spectroscopy has resolved graphene defects and localized states~\cite{Rutter2007,Ugeda2010,Mao2016}. For weak tunneling, low temperature and comparable tip matrix elements on the two sites,
\[
 \frac{(dI/dV)_A-(dI/dV)_B}{(dI/dV)_A+(dI/dV)_B}
 \simeq\frac{\nu_A(E,\boldsymbol x)-\nu_B(E,\boldsymbol x)}
 {\nu_A(E,\boldsymbol x)+\nu_B(E,\boldsymbol x)}.
\]
Here $E=E_F+eV$, and $\nu_A$ and $\nu_B$ are sublattice-resolved local densities of states, requiring a spectral sum over normalized states from both incidence directions and all angular and valley sectors. In the present ideal model, equal spectral weights of opposite angular channels imply $\nu_A=\nu_B$. The channel-resolved profiles in Fig.~\ref{fig:pseudospin} cancel in the equilibrium sublattice difference under these symmetric weights. The full spectral sum is needed to calculate the geometry-dependent total local density of states and a quantitative tunneling spectrum. An occupation imbalance between the relevant sectors or a symmetry-breaking perturbation is necessary for a nonequilibrium sublattice signal.

A possible extension is to replace the ring-gate potential by a molecule--graphene interaction. Theoretical studies of graphene nanoribbons with nanopores predict nucleobase-dependent conductance changes~\cite{Nelson2010,Saha2012}, while calculations for graphene quantum point contacts show that geometry and gate-controlled carrier concentration can tune sensitivity to molecular charges~\cite{Girdhar2013}. DNA passage through graphene nanopores has been demonstrated experimentally through ionic-current changes~\cite{Schneider2010}; this is distinct from electronic transport along a curved graphene junction.

In such a device, a molecule would pass through the throat while the electronic current flows along the graphene surface between the two layers. Comparing $\Delta G=G_{\rm molecule}-G_{\rm empty}$ between symmetric and asymmetric junctions requires the same molecular conditions in both. The molecule--graphene interaction would replace the prescribed Gaussian potential and generally mix angular channels. A calculation of detection or base discrimination must include molecular position and orientation, electrolyte screening, finite temperature and measurement noise. Experimental implementation also requires controlled fabrication of the junction and molecular translocation. The present calculation does not quantify molecular sensitivity or base selectivity.

\section{\label{sec:limits}Consistency Checks and Scope of the Model}

\emph{Large-radius limit ($a\to\infty$, fixed $r_{\mathrm L},r_{\mathrm R}$).}---The throat length obeys $u_p-u_m\sim(r_{\mathrm R}^2+r_{\mathrm L}^2)/a$, and the angular term vanishes for fixed $m$. The matching radii grow as $a$, so the large-argument Hankel asymptotes give $|T_\alpha|^2\to1$ at fixed positive energy and fixed angular channel. The curved interval shrinks at finite interior curvature $K=-1/r_j^2$; there is no pointwise flattening within the throat. The limit is also not uniform over unbounded $m$.

\emph{Zero-length throat ($r_{\mathrm L},r_{\mathrm R}\to0$, fixed $a$).}---The mouth coordinates tend to zero and $R(u)$ tends to $a+|u|$. In the rescaled first-order system, the coefficients are bounded and the propagation interval vanishes. The throat propagation matrix tends to the identity; the individual basis matrices need not have that limit. Consequently $\mathbf W\to\mathbf M_{\rm top}^{-1}\mathbf M_{\rm bot}$, evaluated at $R=a$. A cusp remains in the limiting profile, with $R'$ jumping from $-1$ to $+1$, and the two Hankel orders generally differ. Finite boundary reflection can therefore remain. This limiting junction is not smooth.

\emph{Symmetric profile ($\eta=1$).}---Setting $r_{\mathrm L}=r_{\mathrm R}=r$ makes the two regional hypergeometric bases identical at $u=0$, so their internal matching factors cancel exactly. This recovers a single symmetric hyperbolic throat. It does not convert the proper-distance radius into that of a minimal catenoid.

\emph{Numerical checks.}---We checked six parameter sets spanning low and high transmission, symmetric and asymmetric profiles, and both signs of $m$. Analytic matching at 60 and 100 decimal digits agreed after conversion to double precision. Independent integration of the rescaled first-order system with a high-order adaptive solver agreed with the complex transmission and reflection amplitudes to within $3\times10^{-12}$ in absolute value. The two basis functions also satisfy Eq.~\eqref{eq:masterode} under direct numerical differentiation, with relative residuals below $10^{-40}$ at the tested interior points.

At these six parameter sets, decreasing $\Delta E$ from $10^{-5}$ to $10^{-6}$~eV changed $\tau_W$ by less than $5\times10^{-7}$~fs. These error estimates refer to the sampled parameter sets. Flux conservation is checked at every production evaluation. The local profiles were separately checked against $RJ^u=-4\vF|T|^2/(\pi k)$ across both mouths and the central interface. At $E=300$~meV, tightening the angular-pair tolerance from $10^{-7}$ to $10^{-10}$ changed $G/G_0$ by less than $10^{-7}$ for each plotted asymmetry.

\emph{Scope of the response calculation.}---The prescribed metric, angular boundary condition and scalar potential specify the scattering problem. We obtain the conductance response from potential derivatives of the scattering amplitudes~\cite{Gorini2013} and resolve its energy--position bands into angular-channel contributions. A single-channel density or local density of states alone does not determine these derivatives. For molecular perturbations, the interaction and measurement model enters as discussed in Sec.~\ref{sec:gate}.

High transmission alone does not establish Klein tunneling, and the resonance and bound-state spectra remain to be determined by a separate spectral analysis. A Hartman regime would have to be established through barrier-length dependence and wave-packet propagation~\cite{Hartman1962,Buttiker1982}; negative phase times or weak length dependence of the phase time are insufficient.

\section{\label{sec:conclusions}Discussion and Conclusions}

A scalar ring gate distinguishes the two orientations of an asymmetric graphene throat through their conductance response. At $E=200$~meV and a fixed gate position $u_0=7$~nm, mirror configurations with the same $G_{\rm empty}/G_0=3.08066$ respond to $U_0=+1$~meV with conductance changes of $+0.0718\%$ and $-0.0918\%$. The sign contrast also persists in the comparison at fixed throat length. Displaced gates inside both curved segments also give a contrast in magnitude; the opposite-sign responses occur near the mouths and farther out in the sampled domain. The response resolves spatial differences that the unperturbed two-terminal conductance leaves undetermined.

 Interchanging the upper and lower curvature scales leaves the total Landauer conductance unchanged at fixed minimum radius. For incidence from the upper layer, the local sublattice densities differ between the two configurations. In the example at $E=150$~meV, the mirror pair differs from the symmetric conductance by only $0.073\%$, whereas the most negative sublattice density difference changes from approximately $-0.131$ to $-0.201$ between the mirror profiles under a common incident normalization. The nearly coincident conductance curves accompany appreciable changes in individual-channel densities, whose relative sublattice weights are measured by the normalized polarization.

Higher angular channels are strongly suppressed at low energy by the radius-dependent angular term in the massless Hamiltonian; their contribution to transport grows as the energy increases. The local sublattice imbalance is a channel-resolved property; it cancels, together with the azimuthal current, for equally occupied opposite angular channels. The spin connection remains part of the kinetic operator. The transmission phase derivatives also vary between profiles. Their conversion to time delays depends on the asymptotic phase reference.

Simultaneous reflection of the geometry and gate position leaves the response unchanged. Asymmetry shifts the location and energy of the strongest response; at fixed curvature-scale sum, the maximum magnitude for the asymmetric pair is about $1.5\%$ below the symmetric value in the sampled domain. Hankel and hypergeometric matching preserves radial flux, and independent integration of the first-order equations verifies the unperturbed amplitudes. Variational equations give the gate derivative, with finite-potential solutions checking the linear approximation.

Conductance measurements probe transmission between the layers, and local spectroscopy probes the total local density of states. A spatially controlled gate offers a direct test of the calculated response. For molecular sensing, the gate potential must be replaced by a microscopic molecule--graphene interaction, with the measurement model discussed in Sec.~\ref{sec:gate}.

\section*{Statements and Declarations}
\paragraph{Funding.} This work was partially supported by the National Scientific and Technical Research Council (CONICET) and the Universidad Nacional del Sur (UNS). A.G., F.A., and J.S.A. acknowledge support as members of both institutions.

\paragraph{Competing interests.} The authors declare no competing interests.

\paragraph{Data and code availability.} The numerical codes and datasets generated during the current study are available from the corresponding author on reasonable request.

\appendix
\section{\label{app:hypergeometric}Derivation of the Hypergeometric Throat Solutions}

For the piecewise asymmetric geometry, this derivation holds locally in each domain by substituting $r$ with $r_R$ or $r_L$. We drop the regional labels here for clarity. To solve Eq.~\eqref{eq:masterode} inside the throat, we first use $X\equiv\sinh(u/r_j)$ (satisfying $\cosh^2(u/r_j)-\sinh^2(u/r_j)=1$). The longitudinal derivative becomes $\partial_u = \frac{\sqrt{X^2+1}}{r_j}\partial_X$. Equation~\eqref{eq:masterode} becomes
\begin{equation}
 (X^2\!+1)\,\varphi_{XX} + 2X\,\varphi_X
 + \!\left[k^2r_j^2 + \tfrac{1}{4}
  + \frac{C_0+C_1 X}{X^2+1}
 \right]\varphi = 0.
 \label{eq:throatode_app}
\end{equation} 
The substitution $Y=iX$ converts Eq.~\eqref{eq:throatode_app} to the form
\begin{equation}
 (1{-}Y^2)\varphi_{YY} - 2Y\varphi_Y
 - \!\left[A + \frac{C_0 - iC_1 Y}{1-Y^2}
 \right]\varphi = 0,
 \label{eq:Yode_app}
\end{equation}
with $A\equiv k^2r_j^2+\tfrac{1}{4}$, $C_0\equiv\tfrac{1}{4}-\tfrac{m^2r_j^2}{a^2}$, and $C_1\equiv\tfrac{\alpha m r_j}{a}$. Because the manifold is piecewise, this differential equation is solved locally in each region, where the generic curvature parameter $r_j$ takes the local value $\rR$ for the upper throat ($u > 0$) and $\rL$ for the lower throat ($u < 0$). This equation has regular singular points at $Y=\pm1$ and $Y=\infty$.

Writing Eq.~\eqref{eq:Yode_app} near $Y=-1$ (setting $\xi=1+Y\to0$), the Frobenius indicial equation yields exponents $\pm\gamma_1$, while near $Y=+1$ it yields $\pm\gamma_2$, as defined below Eq.~\eqref{eq:phithroat}.
To remove the singularities, we substitute $\varphi=(1+Y)^{\gamma_1}(1-Y)^{\gamma_2}\Phi(Y)$. A direct calculation yields the Jacobi equation for $\Phi$
\begin{multline}
 (1-Y^2)\Phi_{YY}
 + 2\!\left[\gamma_1-\gamma_2-(\gamma_1+\gamma_2+1)Y\right]\!\Phi_Y \\
 - \left[(\gamma_1+\gamma_2)(\gamma_1+\gamma_2+1)+A\right]\!\Phi = 0.
 \label{eq:Jacobi_app}
\end{multline}
Finally, mapping $z=(1-Y)/2\in\mathbb{C}$ (using consistent branch choices for the complex plane in both the $u>0$ and $u<0$ domains before matching at $u=0$), we obtain the standard Gauss hypergeometric equation
\begin{equation}
 z(1\!-\!z)\Phi_{zz}
 + \bigl[c_h-(a_h+b_h+1)z\bigr]\Phi_z
 - a_h b_h\,\Phi = 0,
 \label{eq:hypergeom_app}
\end{equation}
where the parameter mappings $c_h = 2\gamma_2+1$, $a_h+b_h = 2(\gamma_1+\gamma_2)+1$, and $a_h b_h = (\gamma_1+\gamma_2+\tfrac{1}{2})^2+k^2r_j^2$ lead directly to the solutions presented in Eq.~\eqref{eq:phithroat} and the subsequent text.

\section{\label{app:gate}Linear Conductance Response to a Scalar Gate}

For the rescaled two-component spinor $\boldsymbol f_m=\sqrt{R}\,\psi_{\alpha,m}$, Eq.~\eqref{eq:gate_potential} gives
\begin{equation}
 \begin{split}
 \boldsymbol f_m'&=\mathsf A_m(U_0;u)\boldsymbol f_m,\\
 \mathsf A_m&=\frac{\alpha m}{R}\sigma_3+
       i\frac{E-U_0g}{\hbarvF}\sigma_1.
 \end{split}
 \label{eq:gate_radial}
\end{equation}
Here $\mathsf A_m$ generates spatial propagation, and primes denote $u$ derivatives. The Hankel bases are unchanged at infinity, where the potential vanishes. Starting from a unit outgoing coefficient on the lower layer, propagation and projection onto the upper-layer basis give the incident coefficient $c_m(U_0)$ and hence $T_m=1/c_m$.

Define $\boldsymbol q_m=\partial_{U_0}\boldsymbol f_m|_0$. Differentiation at fixed $E$, geometry, $u_0$ and $w$ gives
\begin{equation}
 \boldsymbol q_m'=\mathsf A_m(0;u)\boldsymbol q_m
       -\frac{i g(u;u_0,w)}{\hbarvF}\sigma_1\boldsymbol f_m^{(0)}.
 \label{eq:gate_variational}
\end{equation}
Here $\boldsymbol f_m^{(0)}=\boldsymbol f_m|_0$. The initial value of $\boldsymbol q_m$ is zero because the outgoing boundary coefficient is held fixed. Projection onto the upper-layer basis gives $c_m=c_m(0)$ and $d_m=\partial_{U_0}c_m|_0$, so
\begin{align}
 \left.\partial_{U_0}T_m\right|_0&=-\frac{d_m}{c_m^2},\nonumber\\
 \chi_G&=-2\sum_m |T_m|^2\operatorname{Re}\frac{d_m}{c_m}.
 \label{eq:gate_derivative}
\end{align}
The sum includes both signs of $m$, with equal transmission probabilities and equal derivatives of those probabilities. Spin and valley degeneracy are included in $G_0=4e^2/h$. Integration of Eq.~\eqref{eq:gate_variational} gives the derivative at zero potential; separate integrations at finite $U_0$ test the linear approximation.

We integrate with DOP853, relative tolerance $2\times10^{-11}$, absolute tolerance $2\times10^{-13}$ and maximum step $w/2$, splitting the propagation at both mouths and the waist. The endpoints lie in the flat layers at least $8w$ beyond the extreme gate centers. Before truncation, three successive angular pairs must each contribute less than $10^{-10}$ to $G/G_0$ and $10^{-9}$~eV$^{-1}$ to every sampled response. Each map point is evaluated directly at the grid spacings stated in Sec.~\ref{sec:gate}.

Unperturbed amplitudes agree with analytic matching within $10^{-12}$ in 48 comparisons. At 54 energy--position--geometry combinations, extrapolated centered differences agree with the variational derivative within $10^{-9}$~eV$^{-1}$. Tighter tolerances, longer tails and additional angular pairs change the tested responses by less than $6\times10^{-11}$~eV$^{-1}$. An interlaced-grid check at 6,300 points gives an interpolation error below $0.031\%$ of the corresponding map maximum at the sampled points; it does not bound relative errors near zeros of the response.

\end{document}